\documentclass[10pt,preprint2]{aastex}
\newcommand{\xt}     {\mbox{\rm\,XTE~J1739--302}}
\newcommand{\dg}        {\mbox{$^{\circ}$}}
\newcommand{\onee}  {\mbox{\rm\,1E~1740.7--2942}}

\begin{document}

\title{XTE~J1739--302 as a Supergiant Fast X-ray Transient}

\author{D. M. Smith\altaffilmark{1}, W. A. Heindl\altaffilmark{2},
C. B. Markwardt\altaffilmark{3}, J. H. Swank\altaffilmark{3},
I. Negueruela\altaffilmark{4}, T. E. Harrison\altaffilmark{5},
and L. Huss\altaffilmark{6}}

\altaffiltext{1}{Physics Department and Santa Cruz Institute for Particle Physics,
University of California, Santa Cruz, 1156 High St., Santa Cruz, CA 95064}
\altaffiltext{2}{Center for Astrophysics and Space Science, University of 
California, San Diego, La Jolla, CA 92093}
\altaffiltext{3}{NASA's Goddard Space Flight Center, Code 662, Greenbelt, MD, 20771}
\altaffiltext{4}{University of Alicante, Apartado de Correos 99, E03080 Alicante, Spain}
\altaffiltext{5}{Astronomy Department, New Mexico State University, Box 30001/Dept. 4500, Las Cruces, NM 88003} 
\altaffiltext{6}{California State University, Hayward, 25800 Carlos Bee Boulevard, 
Hayward, CA 94542}

\begin{sloppypar}

\begin{abstract}

\xt\ is a transient X-ray source with unusually short outbursts,
lasting on the order of hours.  Here we give a summary of X-ray
observations we have made of this object in outburst with the {\it
Rossi X-ray Timing Explorer ({\it RXTE})} and at a low level of
activity with the {\it Chandra X-ray Observatory}, as well as
observations made by other groups.  Visible and infrared spectroscopy
of the mass donor of \xt\ are presented in a companion paper.  The
X-ray spectrum is hard both at low levels and in outburst, but
somewhat variable, and there is strong variability in the absorption
column from one outburst to another.  Although no pulsation has been
observed, the outburst data from multiple observatories show a
characteristic timescale for variability on the order of 1500--2000~s.
The {\it Chandra} localization (right ascension $17^{\rm h} 39^{\rm m}
11.58^{\rm s}$, declination $-30\dg ~20'~37.6''$, J2000) shows that
despite being located less than 2\dg\ from the Galactic Center and
highly absorbed, \xt\ is actually a foreground object with a bright
optical counterpart.  The combination of a very short outburst
timescale and a supergiant companion is shared with several other
recently-discovered systems, forming a class we designate as
Supergiant Fast X-ray Transients (SFXTs).  Three
persistently bright X-ray binaries with similar supergiant companions have also
produced extremely short, bright outbursts: Cyg~X--1,
Vela~X--1, and 1E~1145.1--6141.

\end{abstract}

\section{Introduction}

Most bright X-ray transients last for weeks, and come from two kinds of
systems: low-mass X-ray binaries where the compact object is either a black
hole or a neutron star with a low magnetic field (X-ray novae), and
neutron star binaries with a high-mass, Be-type companion (Be/NS).  
In the latter systems, the neutron star is usually in a wide,
eccentric orbit, and the outbursts often occur at regular intervals, separated
by the orbital period, and generally close to periastron.
These two classes can be distinguished from each
other by neutron star pulsations, by optical observations
of the companion to determine the compact object's mass, by regularity of
recurrence (which indicates a Be/NS binary), or by their X-ray
spectra.

On 12 August 1997, a new X-ray transient near the Galactic Center,
\xt, was discovered and localized using two serendipitous scans by the
{\it Rossi X-ray Timing Explorer ({\it RXTE})}, about 2.5~hr apart and
in nearly perpendicular directions \citep{Sm98b}.  Null results from
other {\it RXTE} exposures to the field limited its duration to less
than 11~dy.  Although this is unusually short for a Be/NS binary
outburst, and although no pulsations were detected despite a very high
sensitivity to periods up to 300~s, we concluded that the system was
probably a Be/NS system with a long spin period and short outburst
duration.  This conclusion was based primarily on the X-ray spectrum,
which was well fitted by an optically thin thermal bremsstrahlung
spectrum of temperature (21.6~$\pm$~0.8)~keV, a spectral form typical
of X-ray pulsars such as Be/NS systems.  The hydrogen column derived
from the same {\it RXTE} spectrum, (6.2~$\pm$~0.2)$\times
10^{22}$~cm$^{-2}$, was consistent with absorption along the line of
sight to the Galactic Center, reinforcing the conclusion that this was
indeed a source in the Galactic bulge -- in which case it was also
unusual as a Be/NS outburst in its relatively high luminosity.

In this paper, we summarize all the X-ray observations of this system
to date, revealing variations in spectrum, luminosity, and absorption
column.  We discuss the initial identification of the counterpart in
the United States Naval Observatory (USNO) and Two Micron
All Sky Survey (2MASS) catalogs.

In the companion paper \cite[hereafter Paper II]{Ne05a} we discuss the
results of detailed optical and infrared spectroscopy of the companion.
Both of our original conclusions --
that this is a Be/X-ray system and that it is located in the Galactic
bulge -- are incorrect.  The companion
is a highly reddened blue supergiant in the
foreground, and the absorbing material is local to the system (Paper II).
\xt\ is therefore an unusual system in having
a blue supergiant companion and very brief outbursts.
We conclude with a discussion of the connection of this object with
other sources that have similar properties, defining a new
subclass of high-mass X-ray binary system.  We also
discuss the possibility that the same flaring process is active
in three continuously bright binaries that also have short outbursts
and blue supergiant companions: Cyg~X--1, Vela~X--1, and
1E~1145.1--6141.

\section{Observations}

Table~1 shows a summary of observations of outbursts of \xt, along with
the two sensitive observations made by focusing X-ray observatories.
Most of the outbursts discovered so far have been the result of
two long-term studies with {\it RXTE}, neither of them specifically
directed toward this system.  

\subsection{Outbursts from the 1E~1740.7--2942 campaign}

Since the start of {\it RXTE} operations in 1996, we have been
observing the black-hole candidate \onee\ regularly \citep[and
references therein]{Sm02}, starting at monthly intervals and gradually
increasing over the course of the mission to twice per week.  The
field of view of the {\it RXTE} pointed instruments when observing
\onee\ includes \xt, at an off-axis position corresponding to a
response of 20\% relative to the center of the field (because of the
need to avoid other sources in the region, \onee\ itself is placed at
a relative response of 43\%).  \xt\ was first discovered because {\it
RXTE} slewed almost directly across it during an outburst as it
approached the \onee\ field on 12 August 1997.  Just 2.5~hr
previously, a scan that was part of a program to survey the Galactic
plane \citep{Va98} passed over the active \xt\ as well, allowing a
two-dimensional localization to $4.8'$ accuracy (99\% confidence; but
see $\S$2.4 below).

Because \xt\ remained in the field of view during the whole 1300~s of
the pointing toward \onee, this episode also gave us the first good
spectral and timing data from \xt\ \citep{Sm98b}.  We have identified
two other major outbursts in the series of pointings to \onee, in 2000
November and 2002 June.  Figure~1 shows a differential lightcurve of
the PCA count rate in the \onee\ pointings since the start of 1997 --
each point has had the average of its two immediate neighbors
subtracted out to remove the slow variations intrinsic to \onee.  The
count rates shown are from the Proportional Counter Array (PCA)
instrument, in the energy range 2.5--25~keV, normalized to a single
proportional counter unit (PCU) and using the first xenon layer of the
detectors only.  Detector PCU0, which suffered a puncture in the middle of
the mission, degrading its performance, is not used after that
event.  The transient outbursts we report were visible in all PCU units
that were on at the time of observation.
Instrumental background was subtracted using the
``cmfaintL7'' model, which is accurate to about 2\%, \citep{Ja05}; the
typical count rate due to \onee\ is on same order as the background.
Galactic diffuse emission, which is significant below 10~keV, was
estimated by pointing to a blank field nearby \citep{Ma99}.  Since
this emission is constant it is irrelevant to the differential
lightcurve in Figure~1, but it is important for spectroscopy.

Two points stand out: the discovery outburst early on, and a bright
outburst in 2002.  We give the two moderately high points nearer the
middle of the plot two very different interpretations.  The second of
them involves not just an increase in high energy counts in the \onee\
field but also a decrease in low energy counts -- we therefore take it
to be a brief, unusual, temporary change in spectral state of \onee\
\citep{Sm02}.  The earlier of these two points at around 200 c/s/5PCU,
however, is consistent with an additional spectral component added on
top of the average \onee\ spectrum from the previous and subsequent
several pointings.  The lightcurve of this pointing is similar to the
two very bright outburst pointings in that it has significant
low-frequency noise, which is never the case with \onee\ in its hard
state (Figure~2).  If the expected \onee\ contribution is subtracted
and the presumed \xt\ spectrum is fit, it shows extremely high
absorption (Table~1).  The unabsorbed luminosity of this outburst is
then actually comparable to the other two events in this survey, although
its raw count rate is much lower.

Table~2 shows the result of fitting each of these three PCA outbursts
from 2.5--25~keV with the optically thin thermal bremsstrahlung model from
\citet{Ke75} as provided by XSPEC \citep{Ar96}.  \citet{Lu05} found
that the thermal bremsstrahlung curve we fit to the first major
outburst observed by {\it RXTE} \citep{Sm98b} also fit perfectly to the
{\it Integral}/IBIS/ISGRI data from the outbursts on MJD 52877 (26 August 2003)
and MJD 52888 (see Table~1).  Together, these results suggest that the
intrinsic spectra in outburst are relatively consistent,
although the amount of local absorbing material is highly variable.  
Using the distance to \xt\ of 2.3~kpc derived photometrically in Paper II,
the bright outbursts peak at a few $\times 10^{36}$ erg s$^{-1}$ in 
unabsorbed luminosity from 1--200~keV (Table~2).

\subsection{Outbursts from the Galactic bulge scan campaign}

The PCA Galactic bulge scan campaign \citep{Sw01} has taken data
approximately every three days since 5 February 1999.  Although
spectral and fast-variability information are not obtained from these
data, the identification of the outbursts with a particular source is
much more secure, since the instrument's one-degree field of view is
scanned rapidly over the Galactic center region.  Figure~3 shows the
lightcurve for \xt\ from this campaign.  There is only one outburst
whose order of magnitude is the same as the three bright outbursts
found in the \onee\ campaign since 1996.  Since we cannot derive an
absorption column for this event, we do not know if it is bright but
highly absorbed or somewhat fainter than the other three.  The
superior ability of the scanning technique to distinguish between
small outbursts from this source and small changes in \onee\ has
allowed the identification of many much smaller outbursts (Table~1,
section 2).  It is possible that any number of the outbursts from the
bulge scans might have been highly absorbed like the 2000 November
event and therefore much more luminous than their count rate
indicates.  

\subsection{Outbursts from {\it Integral}}

\xt\ was later observed in outburst in Galactic bulge data from the
{\it Integral} satellite \citep{Su03a, Lu05}.  Data products from three
{\it Integral} observing seasons of the Galactic bulge are now publicly
available from the {\it Integral} Science Data Centre (ISDC).  Figure~4
shows lightcurves from the IBIS/ISGRI instrument over these three
seasons, summed from 13--71~keV.  There is one very significant,
multi-point flare in each season, and a number of minor peaks from 5
to 10 counts per second.  \citet{Lu05} show an additional outburst
around MJD 52888.3, with about half the peak flux of the event on MJD
52877 and a shorter duration.  This interval is not available in the
public data set shown in Figure~4.  \citet{Sg05} show much more detailed
lightcurves of the outburst (or series of outbursts) on 53073/4.
The {\it Integral} feature on MJD 52895
in Figure~4 is only a single point with a larger than normal error
bar, so we are not including it in our catalog of outbursts.

Two of the minor outbursts in the {\it RXTE}/PCA bulge scan data come just
after major outbursts measured by {\it Integral}, suggesting that some of
the other ``minor outbursts'' in the {\it RXTE} bulge scans may also be part of
the decay or rise phase of larger events.  

The observation reported by \citet{Su03a} is in the center panel of
Figure~4, and
is the longest in duration and fluence.  The time structure of this
event was shown by \citet{Lu05} and is also shown in Figure~5.
Because {\it Integral}'s coverage is much more continuous during an
observing season than {\it RXTE}'s, it is the {\it Integral} observations that
show us the true characteristic durations of outbursts from \xt, from
about 2~hr to about half a day.

\subsection{{\it Chandra} observation}

We observed \xt\ with the ACIS-I CCD on {\it Chandra} on 15 October
2001, for 5 ksec in Timed Exposure (TE) mode.  The observation was not
made in response to any outburst, but was scheduled at the convenience
of the observatory.  We found about 0.25
counts per second from the only significant point source in the field
of view.  The position of this object is shown in Figure~6.  It falls
within the quoted error circles from {\it ASCA} \citep[90\%
confidence]{Sa02} and {\it Integral} \citep[confidence level not
quoted]{Lu05}.  It falls barely outside the 99\% confidence levels
from {\it RXTE} in the 1997 outburst \citep{Sm98b} and the 2001
outburst.  We note, however, that {\it RXTE} is not an imaging
instrument, nor were the slews we used to produce these localizations
optimized for the purpose, as are those of the Galactic bulge campaign
\citep{Sw01}.  The 99\% confidence limit for 1997 was based on a set
of simulations of the effect of intrinsic variations in the count rate
of the source.  These variations can distort the position derived from
a slew, which also relies on a time-varying count rate.  The 99\%
confidence limit was based on an assumption that the intrinsic source
variation during the slew was no bigger than the variations observed
in 1300~s of subsequent pointing to the source.  If this assumption
was violated, that confidence limit was too restrictive.  The
circle shown for 2001 simply repeats the uncertainty derived from the
1997 outburst.

Although the observed count rate would normally cause moderate pileup
in ACIS--I, the source turned out to be 4$\arcmin$ from the pointing
axis.  The point-spread function at that distance is broad enough to
spread the image over many more pixels than would have been the case
on axis.  Separate spectra taken from the core and wings of the image
were consistent with each other, indicating that there is no problem
with pileup.  The {\it Chandra} spectrum was well fit by a hard power
law (see Table~1).  When we attempted to fit it with the thermal
bremsstrahlung spectrum that matched the {\it RXTE} data in outburst
(Table~2), we found that the limited energy range of {\it Chandra} did
not allow us to constrain the temperature.  The same problem appeared
when the {\it RXTE} data were artificially constrained to the same
energy range (below 10~keV).  But the power law fits in this
restricted range show that the low-level emission was harder than the
outburst emission (Table~1).

In the 2--10~keV band the unabsorbed luminosity is 8$\times 10^{33}$
erg s$^{-1}$ at 2.3~kpc.  Assuming a spectrum identical to the 1997
outburst, the bolometric luminosity in X-rays of all energies would be
2.8 times higher.  Because we cannot see where the {\it Chandra}
spectrum starts to turn over, we do not know if the true bolometric
luminosity was somewhat more or less than this value.  As pointed out
by \citet{Za05}, the state we observed with {\it Chandra} is not
``quiescent'', even though no attempt was made to trigger on an
outburst.  The {\it ASCA} upper limit (see below) is, after all, an
order of magnitude lower than the {\it Chandra} flux; and the
observation by \citet{Za05} of the similar source IGR~J17544--2619 in
quiescence showed a much lower flux and a very soft spectrum, entirely
different from the hard spectra seen by {\it Chandra} and {\it RXTE}
for \xt.
  
The lightcurve, shown in Figure~2, has a root mean square (rms)
variability of 36.5\%, compared to an expectation of 35.2\%
from Poisson noise, using the binning time of 32.41~s shown in
Figure~2.  In the absence of any nonuniformity in the
instrumental response over the course of the observation,
this suggests an intrinsic rms variability of 10\%.  Fourier
analysis of the lightcurve shows no significant pulsations.

The position derived from this {\it Chandra} pointing
was right ascension = $17^{\rm h}
39^{\rm m} 11.58^{\rm s}$, declination = $-30\dg ~20'~37.6''$, J2000
\citep{Sm03a, Sm04a}; it
enabled us to search for the optical counterpart of the system
(see below).

\subsection{Upper limits}

The single upper limit shown in Table~1, from {\it ASCA} \citep{Sa02},
is included because it is the deepest such limit for this source, and
because it occurred in the same observation that fortuitously saw the
start of a major outburst (MJD 51248.3, Table~1)
immediately afterward.  Due to the variable absorption column from
\xt, we cannot use this upper limit to directly place a limit on the
unabsorbed flux.   Although the source was much fainter than when observed by
{\it Chandra}, we cannot say whether this was due to a higher
absorption column or intrinsically fainter emission.  However, by
assuming that the absorption column was the same as during the
outburst immediately following, \citet{Sa02} found the 3$\sigma$ upper
limit for the observed flux shown in Table~1.  Removing the effect of
absorption and placing the source at 2.3~kpc (Paper II), this corresponds
to 7.0$\times 10^{32}$ erg s$^{-1}$ at 2.3~kpc.  

The fact that this source was not discovered prior to 1997 puts implicit
upper limits on the frequency of its outbursts.  Figure~1 suggests a
duty cycle of $\sim$ 1\% for the bright outbursts from most of
the pointings of {\it RXTE} to \onee; the {\it RXTE} bulge scans (Figure~3)
and {\it Integral} (Figure~4) are better able to separate sources,
and suggest that there is a continuous range of outburst
luminosities from the upper limit seen with {\it ASCA} to the low level seen with
{\it Chandra} up to the brightest outbursts observed.

A series of explicit upper limits dating from
1988 to 1996 were reported by \citet{Al98} in a retrospective
analysis of data from the TTM coded-mask X-ray telescope, part of the
{\it Roentgen Astrophysical Observatory} on the Kvant module of the
Mir space station.  These upper limits are quoted as ranging from
9--30~mCrab during 21 different pointing series.  Each series 
consisted of a number of pointings spread over an interval varying
from about a day to a month.  Due to the very short duration of 
outbursts from this source, the upper limits that are summed from
the more widely spread series may not be as meaningful as the ones based
on data taken during a single day.

\section{Discussion}

\subsection{Identification of the mass donor}

The {\it Chandra} position derived for \xt\ is approximately
0.65$\arcsec$ from a bright star in the USNO~B1.0 catalog, 
0596--0585865 (0525--28760590 in the USNO~A2.0 catalog).  The
red and blue magnitudes in the first USNO epoch (R1 and B1)
are 13.16 and 17.32, respectively, and in the second epoch
(R2 and B2) are 12.93 and 16.97.
This star is also a bright 2MASS source, 17391155--3020380, with 
J=8.60, H=7.82, and K=7.43.

Figure~7 is a raw color/magnitude diagram of USNO A2.0 stars
within 1$\arcmin$ of the {\it Chandra} position.  The star
that is consistent with the {\it Chandra} position is marked
with a box; it is a very bright and very red
star for the field.  A statistical analysis of a larger field
(3$\arcmin$ in radius) gives the probability of any USNO A2.0 star
coming within 0.65$\arcsec$ of the {\it Chandra} position by chance as
$2.3 \times 10^{-3}$.   The probability of such a coincidence with a star as 
bright as 0525--28760590 is $3.5 \times 10^{-5}$ and the probability
of coincidence with a star with as red a B-R color as 0525--28760590 is
$3.5 \times 10^{-6}$.

Followup spectroscopy of this star was taken in the visible and in the
infrared by two of us (IN and TEH, respectively), and is reported
in detail in Paper II.  The brightness of the companion demands a
location closer than the Galactic bulge, despite the position of the
source within 2\dg\ of the Galactic Center and its high X-ray
absorption.  Paper II gives a distance estimate of 2.3~kpc based 
on the new spectroscopy and photometry.

\subsection{A characteristic timescale of variability?}

There have been no pulsations reported from any of the outbursts in
Table~1, with very strong upper limits up to a period of 300~s from
the first reported outburst \citep{Sm98b}.  The {\it ASCA} outburst of
1999, caught at its very beginning, showed 2 peaks with a deep valley
between them, separated by approximately 1500~s \citep{Sa02}.  It is
worth noting that all three outburst lightcurves in Figure~2 have a
variation consistent with that period, although it is very close to
the duration of the observation.  Lightcurves of several of 
the {\it Integral} outbursts in \citet{Sg05} show peaks separated by
approximately 2000~s.  Since the duration of the outbursts is only
a few times this period, it is difficult to ascertain
whether there is a true pulsar spin period in this range or
whether this is just a typical timescale of whatever drives the
outburst.  Either a much larger database of events or a serendipitous
observation of a particularly long outburst will probably be
needed to resolve this question.

\subsection{Fast transients and supergiant companions}

\citet{Ya95} discovered a transient source in the Scutum region with
{\it ASCA}, designated AX~1845.0-0433, that turned on abruptly in the
middle of the observing interval, showing violent variations for
several hours until the pointing ended.  This behavior was remarkably
similar to the {\it ASCA} observation of \xt\ several years later
\citep{Sa02}.  Like \xt, this object had a very hard spectrum in
outburst, reminiscent of an X-ray pulsar, but no pulsations were
observed. \citet{Co96} found an O9.5I supergiant within the {\it ASCA}
error circle that they identified as the counterpart.

Two X-ray transients recently discovered with {\it Integral} also have
very fast timescales: IGR~17544--2619 \citep{Su03b} and
IGR~16465--4507 \citep{Lu04}.  Both have blue supergiant companions,
estimated as spectral type O9Ib and B0.5I, respectively
\citep{Pe05,Ne05b}.  The similarities of both to \xt\ in X-ray
behavior and photometry led us to predict a blue supergiant companion
for the latter source \citep{Sm04b}, which we later confirmed
spectroscopically \citep{Ne05b}.  Rough distance estimates based on
photometry have been published for these systems: 3.3, 8.5, and 12.5~kpc
for \xt, IGR~17544--2619, and IGR~16465--4507 respectively
\citep{Sm04b} and 3.6~kpc for AX~1845.0--0433 \citep{Co96} (see Paper II
for the derivation of the improved value of 2.3~kpc for \xt).  These
four systems seem similar enough to classify them as a new subclass of
high-mass X-ray binary, which we designate Supergiant Fast X-ray Transients
(SFXT). \citet{Za04} also pointed out
the growing connection between fast variability and probable
wind accretion.  

The unknown mechanism that drives the fast outbursts may be at work in
other systems as well.  Three persistently bright X-ray binaries with
blue supergiant donors have recently been reported to show extremely
bright outbursts far beyond their normal luminosity, lasting on the
order of hours: pulsars Vela~X--1 \citep{La95,Kr03} and
1E~1145.1--6141 \citep{Bo04}, and Cyg~X--1, the canonical black-hole
binary, which has shown six such events \citep{Go03}.  This suggests
that the mechanism of this sort of fast outburst is not specific to
the nature of the compact object.  It may lie either in a sudden
expulsion of material from the companion (\citet{Za05} discusses the
possible role of a clumpy wind) or an instability particular to the
small accretion disks expected in a wind-accreting binary.  The great
differences in persistent flux between the SFXT sources like \xt\ and
the bright sources that also flare quickly may lie in the size of the
orbit or the level of activity in the companion; from what is known or
deduced about the orbits and winds of these systems, however, there is
not yet any clear correlation with persistent X-ray emission.

The large and variable local absorption in some of the
systems may eventually provide a clue to the outburst
mechanism.  Vela~X--1 shows a similar broad range of high absorption
column values to \xt\ \citep{Na86}.  Cyg~X--1 has a much lower
absorption column (usually below 10$^{22}$ cm$^{-2}$), which varies
over about a factor of four with the orbital phase of the binary
\citep[who also suggest that partial covering by a clumpy wind
provides a better spectral fit than a simple absorption model]{Ki00}.
There have been no published observations of the large outbursts of
Vela~X--1, Cyg~X--1 or 1E~1145.1--6141
using instruments sensitive below 10~keV that can measure
absorption.

Not every fast X-ray transient can be put into the SFXT category.
Flare stars and RS~CVn systems have long been known to produce short
X-ray flares, but these tend to have softer spectra and a nearly
isotropic distribution, demonstrating that they are nearby
\cite[e.g.][]{Py83,Ca99}.  X-ray superbursts can also have durations
comparable to SFXT flares, but they generally occur on neutron stars
that produce ordinary Type~I X-ray bursts and can be recognized by
their characteristic luminosity, lightcurve, and long recurrence time
\citep{Co00}.  There have also been short outbursts from three unusual
X-ray binary systems that do not have the same sort of companions as
the SFXTs: CI~Cam \citep{Sm98a}, V4641~Sgr \citep{Re02}, and A~0538--66
\citep{Wh78}.  The outbursts from the former two systems were highly
super-Eddington, which distinguishes them from the events discussed
here, and were never repeated; the last, a well-known but mysterious
system, is thought to be an extreme short-period Be/X-ray binary, and
has shown the regular, periodic outbursts often seen in that class.

The 4.7~s pulsar AX~J1841.0--0536, discovered by \citet{Ba01} with
{\it ASCA}, showed flaring by a factor of 10 with a rise time of about
1~hr, and the authors proposed that it might be a new member of the
same class as \xt\ and AX~1845.0-0433; {\it Integral} later
detected a flare with a 10-minute risetime that may have come
from the same system \citep{Ro04,Ha04a}.  \citet{Ha04b} used
{\it Chandra} to identify the counterpart of AX~J1841.0--0536, which
they reported as having features of the B spectral class.  They
also observed a weak, double-peaked H$\alpha$ profile, which they
used to identify the companion as a Be star.  This feature can
also appear in interacting OB supergiants, however, and \citet{Ha04b} did
not report any diagnostics of the luminosity class of the system.
SAX~J1818.6--1703 \citep{Za98}, XTE~J1901+014 \citep{Rem02},
and AX~J1749.1--2733 \citep{Gr04} are all 
fast transients whose optical counterpart is unknown, as was
recently pointed out by \citet{Za05}.  Now
that a tentative connection has been made between supergiant
companions and fast variability, detailed study of the unclassified optical
counterparts to all these systems should be a high priority.

{\bf Note to be added in proof:} {\it The recent association of
short, hard gamma-ray bursts with older stellar populations in
galaxies at Z$\sim$0.2 \citep{Fo05}, and the strengthening of the NS/NS or
NS/BH merger hypothesis for these events, encourages
a search for the progenitors of these systems.  Although we are
just beginning to study how numerous the SFXTs might truly be,
they seem to be a good candidate for the parent population.}

\acknowledgments

Analysis of the {\it RXTE} data was supported by NASA grants
NAG5--13576 and NNG04GP41G.  Analysis of the {\it Chandra} data was
supported by NASA grant GO1--2030X, administered by the Smithsonian
Astrophysical Observatory.  Public {\it Integral} data were obtained
from the ISDC.  IN is a researcher of the program {\em Ram\'on y
Cajal}, funded by the Spanish Ministerio de Ciencia y Tecnolog\'{\i}a
and the University of Alicante, with partial support from the
Generalitat Valenciana and the European Regional Development Fund
(ERDF/FEDER).  This research is partially supported by the Spanish
MCyT under grants AYA2002-00814 and ESP-2002-04124-C03-03.

\end{sloppypar}

\onecolumn

\begin{figure}
\plotone{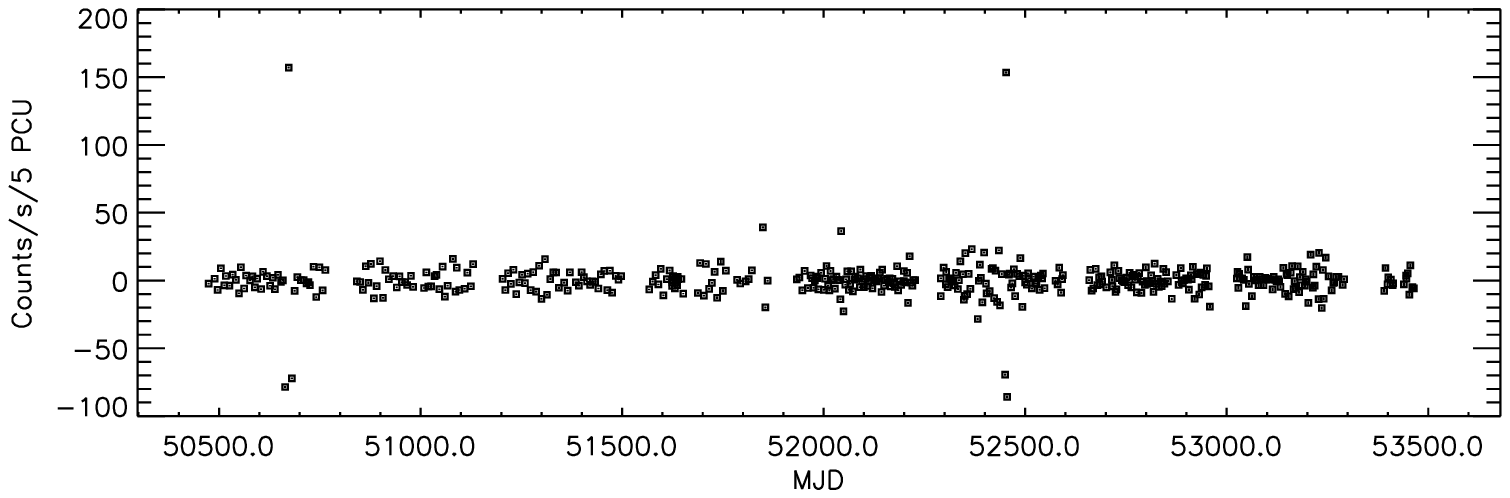}
\caption{
Differential lightcurve for the \onee\ pointings with {\it
RXTE}. The background-subtracted count rate over the full PCA energy range
is shown, scaled to the equivalent value if all 5 PCU detectors were
turned on (typically from one to three actually were), and uncorrected
for the $\sim$ 20\% collimator response due to the pointing direction.
To enhance the visibility of abrupt changes, we subtracted from each
point the average of its neighbors; this also results in the negative points
on either side of the two bright outbursts.  Only 
pointings with another observation occurring within two weeks in both
directions were included.
}
\end{figure}

\begin{figure}
\plotone{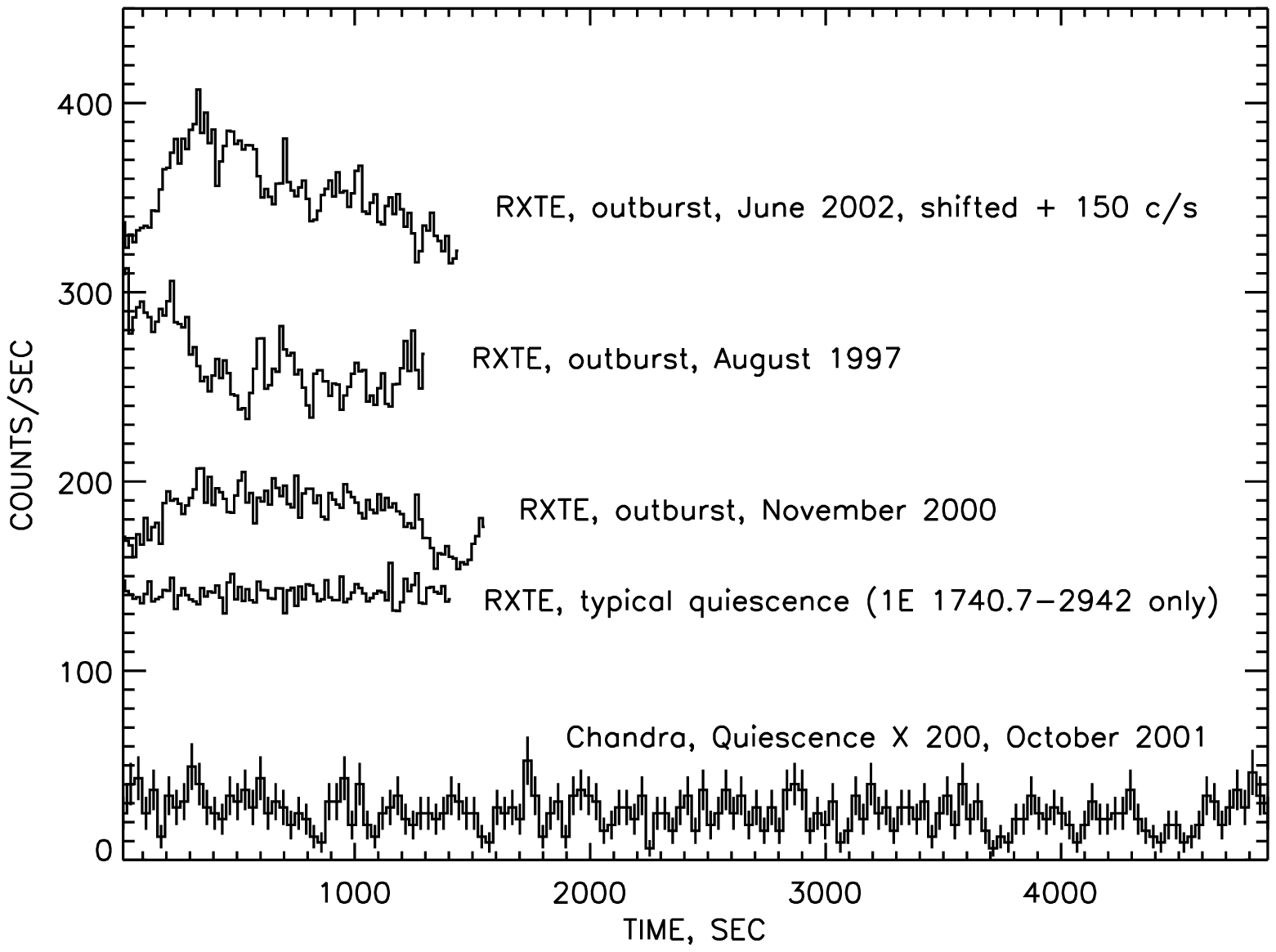}
\caption{
Lightcurves of three individual outbursts of \xt\ from the \onee\
pointings (top three traces, with the flux from \onee\ still present), 
a typical \onee\ lightcurve when
\xt\ is quiescent, and the lightcurve of \xt\ as seen at a low level
of activity with
{\it Chandra}.  The {\it Chandra} trace is multiplied by 200 to be
seen on this scale, and the 2002 outburst has been shifted upward
by 150 counts s$^{-1}$.
}
\end{figure}

\begin{figure}
\plotone{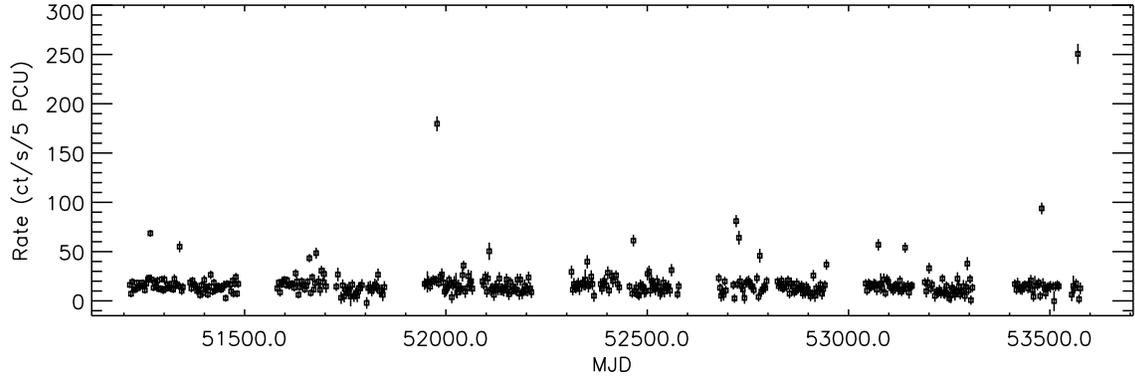}
\caption{
Lightcurve of \xt\ from the {\it RXTE} Galactic bulge scans.  Like
Figure~1, the data shown are equivalent count rates for all 5 PCA
detectors, but here the count rate shown is from 2--10~keV (typically
well over half of the total rate), and the count rates are for 100\%
collimator response (no pointing offset).
}
\end{figure}

\begin{figure}
\plotone{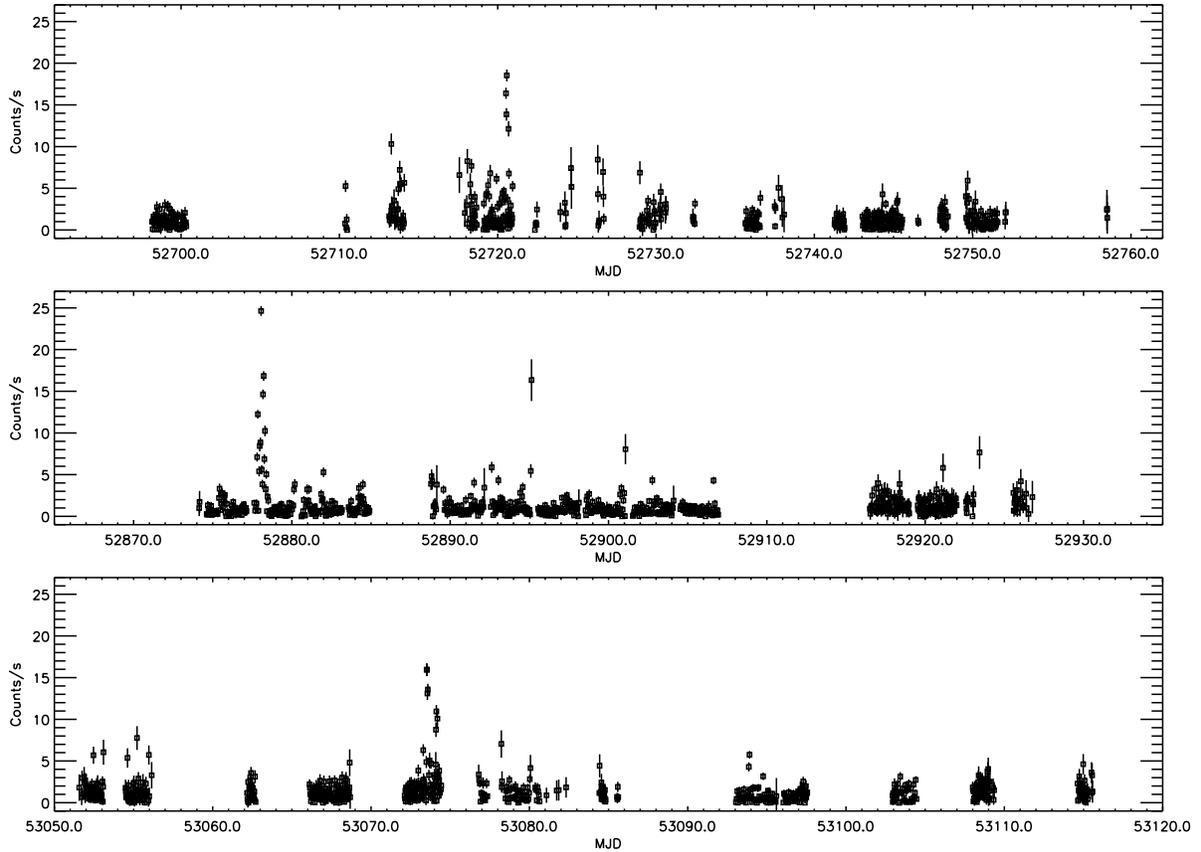}
\caption{
{\it Integral} IBIS/ISGRI lightcurves for \xt\ for three observing seasons,
13--71~keV.  These data are publicly available from the {\it Integral} Science
Data Centre online archive, listed under the source name IGR~J17391--3021.
Five data points with extremely large error bars ($>$ 3 counts per second)
have been removed.
}
\end{figure}

\begin{figure}
\plotone{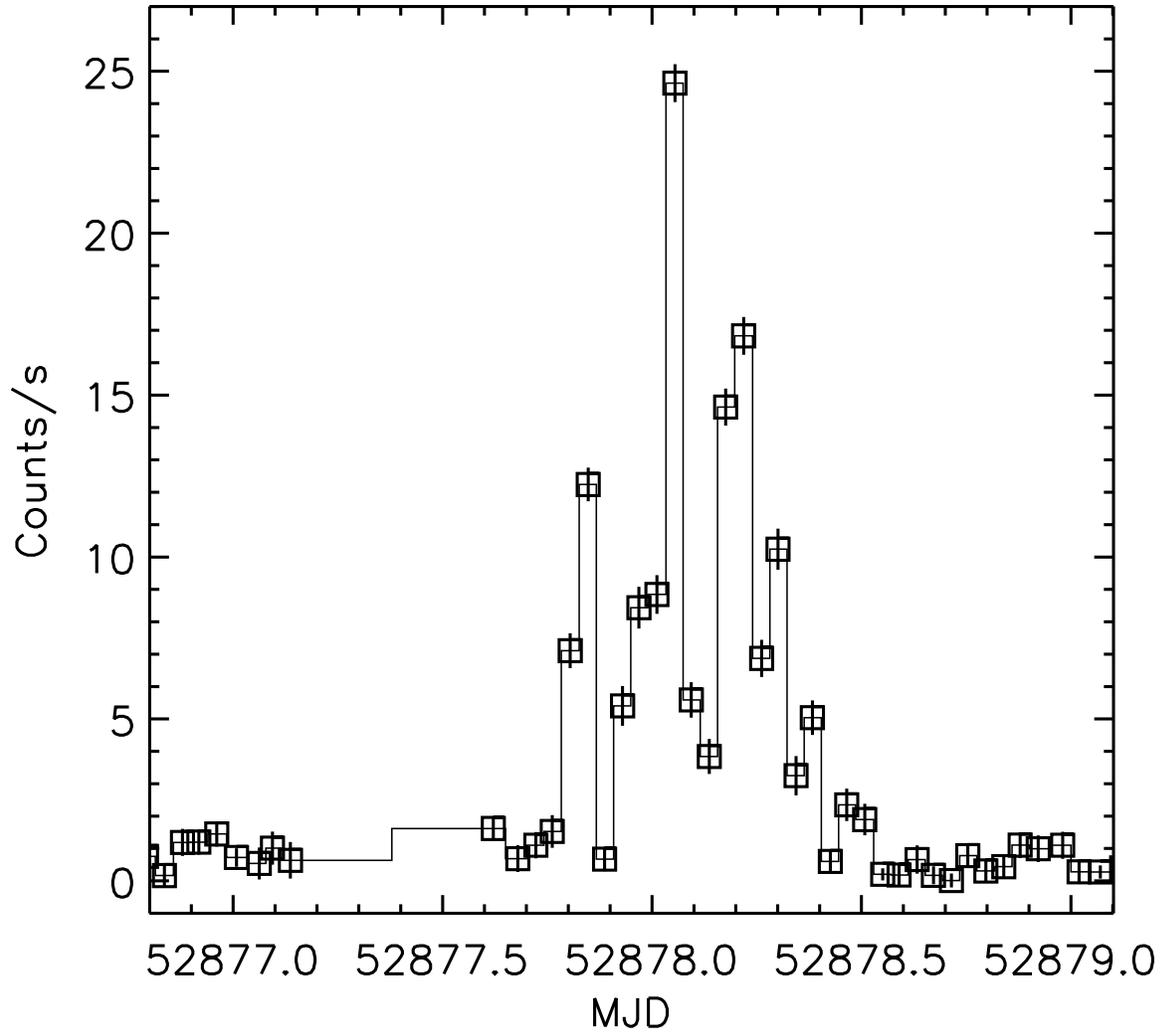}
\caption{
IBIS/ISGRI lightcurve for the most complicated and brightest
of the {\it Integral} outbursts of \xt\ \citep{Su03a}, showing extreme
variability similar to that seen on an even shorter timescale by
{\it ASCA} \citep{Sa02}.
}
\end{figure}

\begin{figure}
\plotone{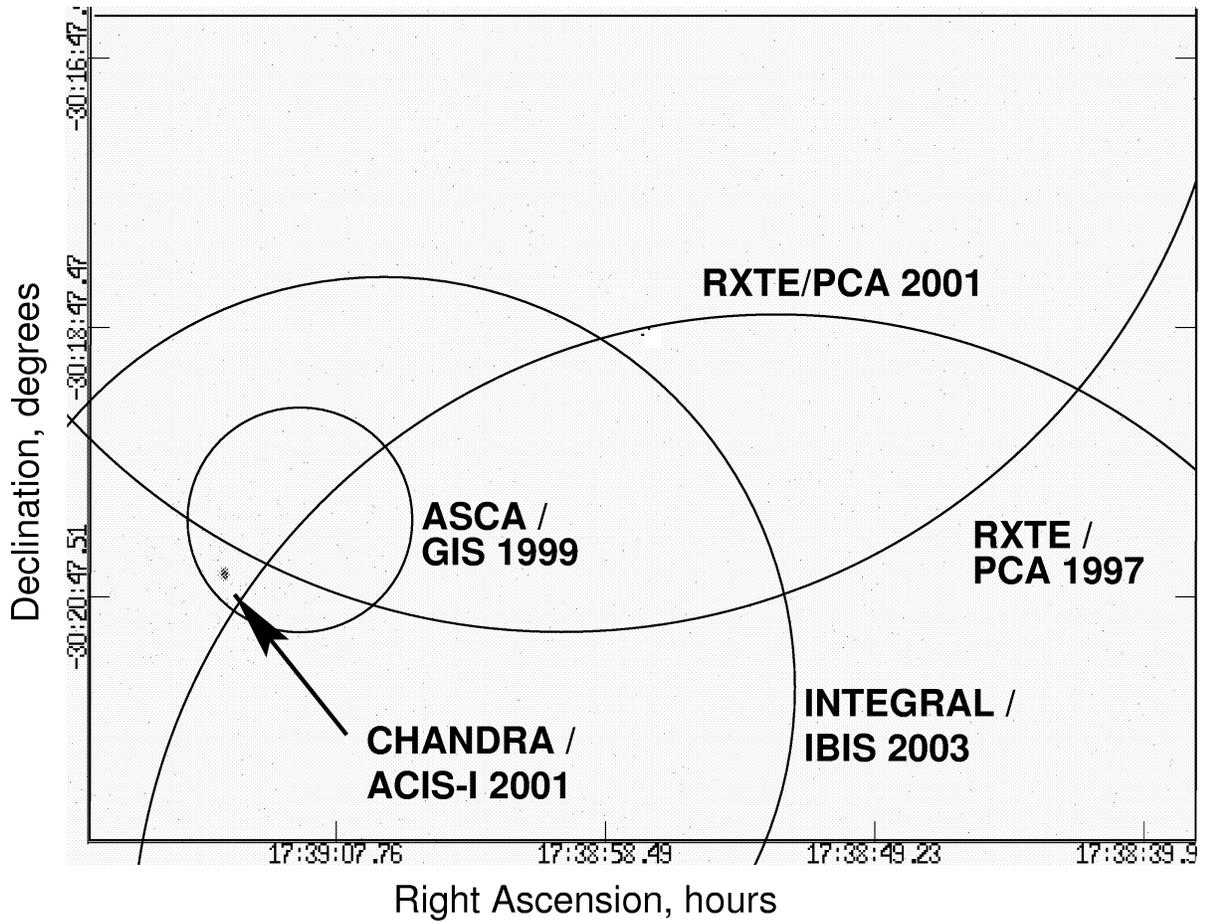}
\caption{
Localization circles for \xt\ from {\it RXTE}, {\it ASCA}, and {\it Integral}
superimposed on part of the {\it Chandra} image field.  All the
circles are consistent or nearly consistent with the {\it Chandra}
source (see text).
}
\end{figure}

\begin{figure}
\plotone{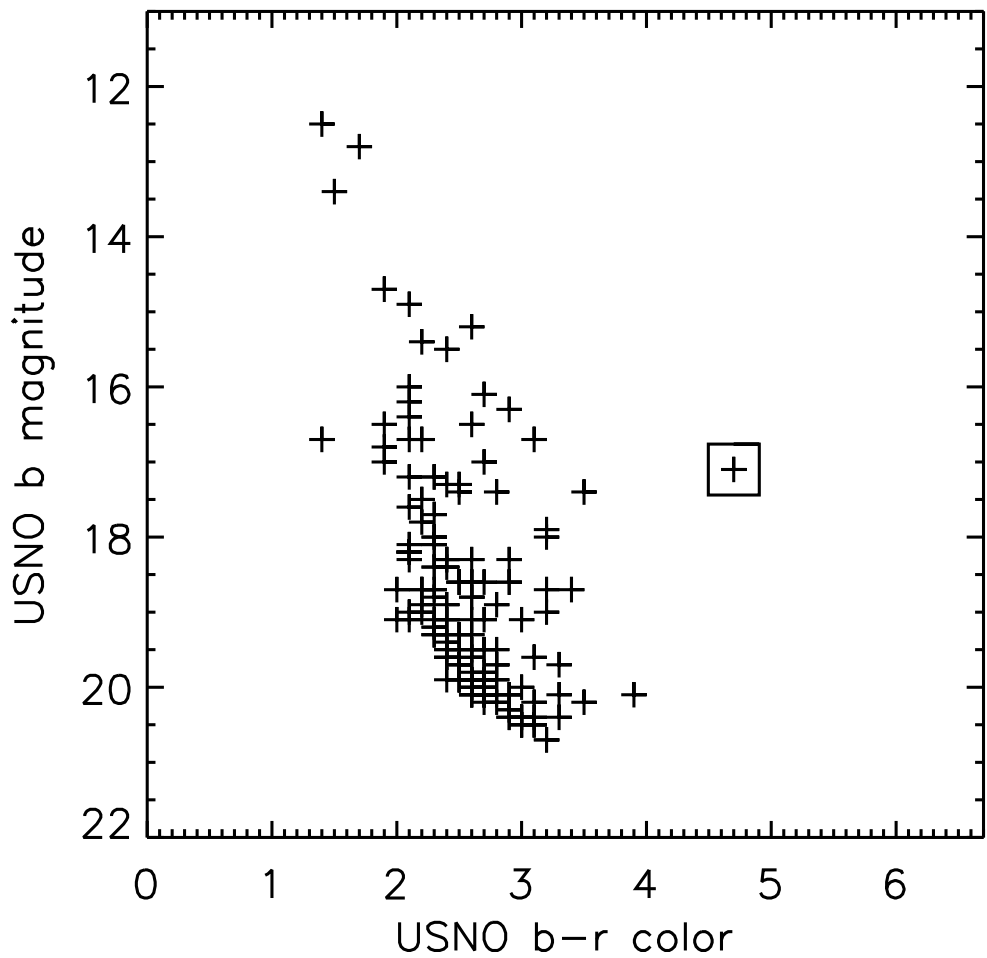}
\caption{
Raw color/magnitude diagram (uncorrected for
absorption) of USNO A2.0 stars within 1$\arcmin$
of \xt.  The square marks the star that
matches the {\it Chandra} position.  
}
\end{figure}

\begin{deluxetable}{llrrrlllll}
\tabletypesize{\scriptsize}
\tablecolumns{7} 
\tablewidth{0pc} 
\tablecaption{Summary of X-ray Observations of XTE~J1739--302}
\tablehead{ 

\colhead{MJD} & \colhead{Instrument} & 
\colhead{PCA} & \colhead{Absorbed} & 
\colhead{Unabsorbed} & \colhead{$n_H$} & 
\colhead{Power} & \colhead{Minimum} & \colhead{Maximum} & \colhead{Ref.\tablenotemark{d}} \\

\colhead{} & \colhead{} & 
\colhead{Rate} & \colhead{Flux\tablenotemark{b}} & 
\colhead{Flux\tablenotemark{b}} & \colhead{$\times 10^{22}$ cm$^{-2}$} & 
\colhead{law} & \colhead{duration} & \colhead{duration} & \colhead{} \\

\colhead{} & \colhead{} & 
\colhead{c/s\tablenotemark{a,b}} & \colhead{$\times 10^{-11}$ } & 
\colhead{$\times 10^{-11}$ } & \colhead{} & 
\colhead{index\tablenotemark{c}} & \colhead{(hr)} & \colhead{(dy)} & \colhead{} \\

\colhead{} & \colhead{} & 
\colhead{} & \colhead{erg/cm$^2$/s} & 
\colhead{erg/cm$^2$/s} & 
\colhead{}  & \colhead{} & 
\colhead{} & \colhead{} & \colhead{} 

}
\startdata
\cutinhead{Major Outbursts:}
50672.7  & {\it {\it RXTE}/}PCA\tablenotemark{e}      & 638 &  169 & 260 & $7.11 \pm 0.31$  &    $1.66 \pm 0.03 $ &  2.5 &  11 & {\it [1]} \\
51248.3  & {\it ASCA/}GIS                     & \nodata &  200 & 236 & $3.17 ^{+0.33}_{-0.31}$ & $0.80 ^{+0.10}_{-0.11}$ & 0.6 & 3.6  & {\it [2]} \\
51849.2  & {\it {\it RXTE}/}PCA\tablenotemark{e}      & 139 &   38 & 155  & $36.8 \pm 4.2$   & $2.04 \pm 0.20$   &  0.3 & 3.9  &   \nodata \\
51977.9  & {\it {\it RXTE}/}PCA\tablenotemark{f}      & 179 & \nodata & \nodata  & \nodata            & \nodata         &      \nodata & 2.9  &   \nodata \\
52452.9  & {\it {\it RXTE}/}PCA\tablenotemark{e}      & 585 &  177 & 257  & $6.37 \pm 0.42$  & $1.55 \pm 0.04$    &  0.3 & 4.9 &    \nodata \\
52720.5  & {\it Integral/}IBIS & \nodata & \nodata & \nodata  & \nodata & \nodata &                  1.8 & 0.08  &    {\it [3]}  \\
52877.8  & {\it Integral/}IBIS & \nodata & \nodata & \nodata  & \nodata & \nodata &                  14 & 0.6  &   {\it [4]} \\
52888.3  & {\it Integral/}IBIS & \nodata & \nodata & \nodata  & \nodata & \nodata &                  7 & 0.3  &   {\it [5]} \\
53073.28\tablenotemark{g}  & {\it Integral/}IBIS & \nodata & \nodata & \nodata  & \nodata & \nodata &                 0.5 & 0.02  &  {\it [3]}    \\
53073.48\tablenotemark{g}  & {\it Integral/}IBIS & \nodata & \nodata & \nodata  & \nodata & \nodata &                  2 & 0.12  &  {\it [3]}    \\
53074.08\tablenotemark{g}  & {\it Integral/}IBIS & \nodata & \nodata & \nodata  & \nodata & \nodata &                  1.3 & 0.05  &  {\it [3]}    \\
53569.7  & {\it {\it RXTE}/}PCA\tablenotemark{f}      & 251 & \nodata & \nodata  & \nodata            & \nodata         &      \nodata & 3.0  &   \nodata \\
\cutinhead{Minor Outbursts:}

51265.80  & {\it {\it RXTE}/}PCA\tablenotemark{f}      &  67  & \nodata & \nodata &  \nodata  &        \nodata   &             \nodata & 7.0 & \nodata \\
51338.45  & {\it {\it RXTE}/}PCA\tablenotemark{f}      &  54  & \nodata & \nodata &  \nodata  &        \nodata   &             \nodata & 7.3 & \nodata \\
51660.76  & {\it {\it RXTE}/}PCA\tablenotemark{f}      &  41  & \nodata & \nodata &  \nodata  &        \nodata   &             \nodata & 7.8 & \nodata \\
51677.68  & {\it {\it RXTE}/}PCA\tablenotemark{f}      &  47  & \nodata & \nodata &  \nodata  &        \nodata   &             \nodata & 8.9 & \nodata \\
52043.77  & {\it {\it RXTE}/}PCA\tablenotemark{f}      &  34  & \nodata & \nodata &  \nodata  &        \nodata   &             \nodata & 6.9 & \nodata \\
52107.51  & {\it {\it RXTE}/}PCA\tablenotemark{f}      &  49  & \nodata & \nodata &  \nodata  &        \nodata   &             \nodata & 6.9 & \nodata \\
52350.98  & {\it {\it RXTE}/}PCA\tablenotemark{f}      &  38  & \nodata & \nodata &  \nodata  &        \nodata   &             \nodata & 8.0 & \nodata \\
52465.63  & {\it {\it RXTE}/}PCA\tablenotemark{f}      &  60  & \nodata & \nodata &  \nodata  &        \nodata   &             \nodata & 5.9 & \nodata \\
52720.90  & {\it {\it RXTE}/}PCA\tablenotemark{f,h}   &  80  & \nodata & \nodata &  \nodata  &        \nodata   &             \nodata & 7.0 & \nodata \\
52727.87  & {\it {\it RXTE}/}PCA\tablenotemark{f}      &  63  & \nodata & \nodata &  \nodata  &        \nodata   &             \nodata & 6.8 & \nodata \\
52779.67  & {\it {\it RXTE}/}PCA\tablenotemark{f}      &  45  & \nodata & \nodata &  \nodata  &        \nodata   &             \nodata & 6.8 & \nodata \\
52945.06  & {\it {\it RXTE}/}PCA\tablenotemark{f}      &  37  & \nodata & \nodata &  \nodata  &        \nodata   &             \nodata & 6.8 & \nodata \\
53074.05  & {\it {\it RXTE}/}PCA\tablenotemark{f,h}      &  57  & \nodata & \nodata &  \nodata  &        \nodata   &             \nodata & 6.8 & \nodata \\
53140.51  & {\it {\it RXTE}/}PCA\tablenotemark{f}      &  54  & \nodata & \nodata &  \nodata  &        \nodata   &             \nodata & 6.8 & \nodata \\
53200.14  & {\it {\it RXTE}/}PCA\tablenotemark{f}      &  33  & \nodata & \nodata &  \nodata  &        \nodata   &             \nodata & 6.8 & \nodata \\
53295.11  & {\it {\it RXTE}/}PCA\tablenotemark{f}      &  38  & \nodata & \nodata &  \nodata  &        \nodata   &             \nodata & 6.8 & \nodata \\
53479.79  & {\it {\it RXTE}/}PCA\tablenotemark{f}      &  94  & \nodata & \nodata &  \nodata  &        \nodata   &             \nodata & 6.8 & \nodata \\

\cutinhead{Observations with focusing X-ray telescopes:}
52197.61  & {\it Chandra/}ACIS-I  & \nodata & 1.1 & 1.3 & $4.2 \pm 1.0$ &  $0.62 \pm 0.23$  &   \nodata & \nodata & \nodata \\
51248.16  & {\it ASCA/}GIS\tablenotemark{i}     & \nodata & $<$0.09 & $<$0.11 & \nodata &         \nodata &              \nodata & \nodata &    {\it [2]} \\

\enddata 
\tablenotetext{a}{Counts/s equivalent for all five detectors of the {\it RXTE} PCA, top xenon layer only, 2--10~keV. }

\tablenotetext{b}{The uncertainties on the photon and energy fluxes are generally dominated by instrument 
                  calibration rather than counting statistics and are on the order of 10\%.}
\tablenotetext{c}{From a fit in the energy range 2--10~keV only.}
\tablenotetext{d}{  {\it [1]} Smith et al. (1998b);  {\it [2]} Sakano et al. (2002);  {\it [3]} 
Sguera et al. (2005); {\it [4]} Sunyaev et al. (2003),
                  Smith et al. (2003b), Rupen, Mioduszewski, \& Dhawan (2003), Lutovinov et al. (2005);
                  {\it [5]} Lutovinov et al. (2005).}
\tablenotetext{e}{Detected during periodic {\it RXTE} pointings to the nearby source \onee.}
\tablenotetext{f}{Detected during periodic {\it RXTE} scans of the Galactic Bulge.}
\tablenotetext{g}{These three {\it Integral} outbursts may also be considered part of a single episode of activity 
                  containing a wide dynamic range of variability.}
\tablenotetext{h}{These observations took place just after one of the major outbursts observed by {\it Integral}/IBIS.}
\tablenotetext{i}{The values are 3$\sigma$ upper limits. The energy range for the unabsorbed flux is not specified
                  in Sakano et al. (2002).}

\end{deluxetable} 

\begin{deluxetable}{lrrr}
\tabletypesize{\scriptsize}
\tablecolumns{7} 
\tablewidth{0pc} 
\tablecaption{Optically Thin Thermal Bremsstrahlung Fits, 2.5--25~keV, {\it RXTE} PCA}
\tablehead{ 
\colhead{MJD} & 
\colhead{$n_H$}  & 
\colhead{$kT$} & 
\colhead{bolo. lum.}  \\
\colhead{} & 
\colhead{($\times 10^{22}$ cm$^{-2}$)} & 
\colhead{(keV)} & 
\colhead{($\times 10^{36}$ erg s$^{-1}$)}  \\
}
\startdata
50672.707  & $6.2 \pm 0.2$  & $21.6 \pm 0.8$ & 5.6 \\
51849.233  & $32 \pm 3$ & $15 \pm 2$ & 2.6 \\
52452.930  & $7.9 \pm 0.3$ & $18.6 \pm 0.8$ & 4.9 \\
\enddata 

\end{deluxetable}


\begin{thebibliography}

\bibitem[Aleksandrovich et al.(1998)]{Al98}
Aleksandrovich, N. L., Borozdin, K. N., Emel'Yanov, A. N., Sunyaev, R. A.,
\&  Skinner, G. K. 1998, Astronomy Letters 24, 742

\bibitem[Arnaud (1996)]{Ar96}
Arnaud, K. A. 1996, ASP Conf. Proc. 101, Astronomical Data Analysis Software 
and Systems V, ed. G. Jacoby \& J. Barnes (San Francisco: ASP), 17

\bibitem[Bamba et al.(2001)]{Ba01}
Bamba, A., Yokogawa, J., Ueno, M., Koyama, K., and Yamauchi, S. 2001
PASJ 53, 1179

\bibitem[Bodaghee, Mowlavi, \& Ballet(2004)]{Bo04}
Bodaghee, A., Mowlavi, N., \& Ballet, J. 2004, ATEL \#290

\bibitem[Castro-Tirado et al.(1999)]{Ca99}
Castro-Tirado, A. J., Brandt, S., Lund, N., and Sunyaev, R. 1999,
A\&A 347, 927

\bibitem[Coe et al.(1996)]{Co96}
Coe, M. J., Fabregat, J., Negueruela, I., Roche, P., and Steele, I. A. 1996,
MNRAS 281, 333

\bibitem[Cornelisse et al.(2000)]{Co00}
Cornelisse, R., Heise, J., Kuulkers, E., Verbunt, F., and in't Zand, J. J. M.
2000, A\&A, 357, L21

\bibitem[Fox et al.(2005)]{Fo05}
Fox, D. B. et al. 2005, Nature, in press, astro-ph/0510110

\bibitem[Golenetskii et al.(2003)]{Go03}
Golenetskii, S. et al. 2003 ApJ, 596, 1113

\bibitem[Grebenev(2004)]{Gr04}
Grebenev, S. A. 2004, talk given at the 5th {\it Integral} workshop, M\"{u}nich, February 2004

\bibitem[Halpern \& Gotthelf(2004)]{Ha04a}
Halpern, J. P., \& Gotthelf, E. V. 2004, ATEL \#341

\bibitem[Halpern et al.(2004)]{Ha04b}
Halpern, J. P., Gotthelf, E. V., Helfand, D. J., Gezari, S., and
Wegner, G. A. 2004, ATEL \#289

\bibitem[Jahoda et al.(2005)]{Ja05}
Jahoda, K. et al. 2005, to be submitted to ApJS

\bibitem[Kellogg, Baldwin \& Koch(1975)]{Ke75}
Kellogg, E., Baldwin, J. R., \& Koch, D. 1975, ApJ 199, 299

\bibitem[Kitamoto et al.(2000)]{Ki00}
Kitamoto, S., Egoshi, W., Miyamoto, S., Tsunemi, H., 
Ling, J. C., Wheaton, W. A., \& Paul, B. 2000, ApJ 531, 546

\bibitem[Krivonos et al.(2003)]{Kr03}
Krivonos, R. et al. 2003, ATEL \#211

\bibitem[Laurent et al.(1995)]{La95}
Laurent, P. et al. 1995, A\&A 300, 399

\bibitem[Lutovinov et al.(2004)]{Lu04}
Lutovinov, A., Rodrigues, J., Budtz-Jorgensen, C., Grebenev, S., and
Winkler, C. 2004, ATEL \#329

\bibitem[Lutovinov et al.(2005)]{Lu05} 
Lutovinov, A, Revnivtsev, M., Molkov, S., and Sunyaev, R. 2005, A\&A 430, 997 

\bibitem[Main et al.(1999)]{Ma99}
Main, D. S., Smith, D. M., Heindl, W. A., Swank, J. H., Leventhal, M., Mirabel,
I. F., and Rodr\'{i}guez, L. F. 1999, ApJ 535, 901

\bibitem[Nagase et al.(1986)]{Na86}
Nagase, F., Hayakawa, S., Sato, N., Masai, K., and Inoue, H. 1986,
PASJ, 38, 547

\bibitem[Negueruela et al.(2005a)]{Ne05a} 
Negueruela, I., Smith, D. M., Harrison, T. E., \& Torrej\'{o}n, J. M. 2005a, ApJ, in press

\bibitem[Negueruela et al.(2005b)]{Ne05b} 
Negueruela, I., Smith, D. M., and Chaty, S. 2005b, ATEL \#429

\bibitem[Pellizza, Chaty \& Negueruela(2005)]{Pe05}
Pellizza, L. J., Chaty, S., and Negueruela, I. 2005, in preparation

\bibitem[Pye \& McHardy(1983)]{Py83}
Pye, J. P. and McHardy, I. M. 1983, MNRAS 205, 875

\bibitem[Remillard \& Smith(2002)]{Rem02}
Remillard, R. A., \& Smith, D. A. 2002, IAUC \#7880

\bibitem[Revnitsev et al.(2002)]{Re02}
Revnitsev, M., Gilfanov, M., Churazov, E., and Sunyaev, R. 2002,
A\&A 391, 1013

\bibitem[Rodriguez et al.(2004)]{Ro04}
Rodriguez, J. et al. 2004, ATEL \#340

\bibitem[Rupen, Mioduszewski, and Dhawan(2003)]{Ru03}
Rupen, M. P., Mioduszewski, A. J., and Dhawan, V. 2003, ATEL \#184

\bibitem[Sakano et al.(2002)]{Sa02}
Sakano, M. et al. 2002, ApJS 138, 19

\bibitem[Sguera et al.(2005)]{Sg05}
Sguera, V. et al. 2005, submitted to A\&A (astro-ph/0509018)

\bibitem[D. A. Smith et al.(1998)]{Sm98a}
Smith, D. A., Remillard, R., Swank, J. H., \& Smith, E. 1998, IAUC \#6855

\bibitem[D. M. Smith et al.(1998)]{Sm98b}
Smith, D. M. et al. 1998, ApJ 501, 181

\bibitem[Smith, Heindl \& Swank(2002)]{Sm02}
Smith, D. M., Heindl, W. A., \& Swank, J. H. 2002, ApJ 569:362

\bibitem[Smith et al.(2003a)]{Sm03a}
Smith, D. M., Heindl, W. A., Swank, J. H., Harrison, T. E., and
Negueruela, I. 2003a, ATEL \#182

\bibitem[Smith et al.(2003b)]{Sm03b}
Smith, D. M., Markwardt, C. B, Swank, J. H., and Heindl, W. A. 2003b, 
ATEL \#186

\bibitem[Smith(2004)]{Sm04b}
Smith, D. M. 2004, ATEL \#338

\bibitem[Smith \& Heindl(2004)]{Sm04a}
Smith, D. M. \& Heindl, W. A. 2004, ATEL \#218

\bibitem[Sunyaev et al.(2003a)]{Su03a}
Sunyaev, R. A., Lutovinov, A., Molkov, S., and Deluit, S. 2003a, ATEL \#181

\bibitem[Sunyaev et al.(2003b)]{Su03b}
Sunyaev, R. A., et al. 2003b, ATEL \#190

\bibitem[Swank \& Markwardt(2001)]{Sw01}
Swank, J. \& Markwardt, C.  2001, in ASP Conf. Ser. 251, 
{\it New Century of X-ray Astronomy}, ed. H. Inoue \& H. Kunieda 
(San Francisco: ASP), 94

\bibitem[Valinia \& Marshall(1998)]{Va98}
Valinia, A. and Marshall, F. E. 1998, ApJ 505, 134

\bibitem[White \& Carpenter(1978)]{Wh78}
White, N. E., and Carpenter, G. F. 1978, MNRAS 183, 11

\bibitem[Yamauchi et al.(1995)]{Ya95}
Yamauchi, S. et al. 1995, PASJ 47, 189

\bibitem[in't Zand et al.(1998)]{Za98}
in't Zand, J. J. M. et al. 1998, IAUC \#6840

\bibitem[in't Zand et al.(2004)]{Za04}
in't Zand, J. J. M., Heise, J., Ubertini, P., Bazzano, A., and Markwardt,
C. 2004, {\it Proc. 5th Integral Workshop, ``The Integral Universe,''}
ed. V. Sch\"{o}nfelder, G. Lichti \& C. Winkler, ESA SP-552, p. 427

\bibitem[in't Zand(2005)]{Za05}
in't Zand, J. J. M. 2005, A\&A 441, L1

\end{thebibliography}
\end{document}